\documentclass[conference]{IEEEtran}
\IEEEoverridecommandlockouts
\usepackage{cite}
\usepackage{amsmath,amssymb,amsfonts}
\usepackage{graphicx}
\usepackage{textcomp}

\begin{document}

\title{Khelte Khelte Shikhi: A Proposed HCI Framework for Gamified Interactive Learning with Minecraft in Bangladeshi Education Systems}

\author{
Mohd Ruhul Ameen\textsuperscript{1},
Akif Islam\textsuperscript{2},
Momen Khandokar Ope\textsuperscript{2}
\\[1ex]
\textsuperscript{1}Department of Computer Science, Marshall University, Huntington, WV, USA
\\[-0.5ex]
\textsuperscript{2}Department of Computer Science and Engineering, University of Rajshahi, Bangladesh
\\[1ex]
Email- ameen@marshall.edu, iamakifislam@gmail.com, khandokermomen919@ru.ac.bd
}

\maketitle

\begin{center}
\small Accepted and presented at the 2025 28th International Conference on Computer and Information Technology (ICCIT 2025), Cox's Bazar, Bangladesh. © 2025 IEEE.
\end{center}

\begin{abstract}
Game-based learning shows real promise for engaging students in well-resourced classrooms, but what about the millions who study in schools with far fewer opportunities? We propose a practical framework for bringing Minecraft Education Edition into Bangladesh’s 130,000 schools, where 55\% lack reliable internet, rural areas receive only 12–16 hours of electricity per day, computer access in rural schools is just 8\%, and student–teacher ratios can reach 52:1. Our approach addresses these constraints directly through three deployment tiers: cloud-based multiplayer for urban schools with stable infrastructure (15\%), local-area network (LAN) solutions with optional solar power for semi-urban schools (30\%), and fully offline, turn-based modes using refurbished hardware for rural contexts (55\%). We provide eight curriculum-aligned Minecraft worlds with complete Bangla localization, covering topics from Lalbagh Fort reconstruction to monsoon flood simulation. The interface supports first-time users through progressive complexity, culturally familiar metaphors rooted in local farming and architecture, and accessibility features such as keyboard-only controls and 200\% text scaling. We outline evaluation benchmarks including 15\% learning gains, 70\% transfer-task mastery, System Usability Scale scores above 70, and a per-student cost below two dollars per hour. Although not yet empirically validated, this work synthesizes game-based learning theory, HCI principles, and contextual analysis to offer implementable specifications for pilot testing in resource-constrained settings. It is presented as a design-oriented conceptual framework rather than a field deployment, providing an implementation-ready blueprint for future empirical validation.
\end{abstract}

\vspace{0.3em}

\section{Introduction}

\subsection{Global Shifts in Education and the Role of Game-Based Learning}
Education is evolving rapidly worldwide. In today’s dynamic environment, students must develop creativity, collaboration, and problem-solving skills, not just memorize facts.\cite{roshid_2024}. In many developing countries, schools strive to prepare young minds for an unpredictable future despite overcrowded classrooms, outdated teaching methods, and fragile infrastructure. Game-based learning offers a promising alternative that enables active engagement, collaboration, and deeper understanding in place of passive memorization \cite{slattery2025_gamebased, alkhyat2023_gbl_effectiveness}. Minecraft stands out not only because millions play it worldwide, but also because recent empirical studies show it can support learning by improving spatial reasoning, engagement, and creative problem-solving in classroom settings \cite{slattery2024_minecraft_study}.

\subsection{Why Bangladesh is a Critical Test Case}
Bangladesh is one of the most populous countries in South Asia, with a population around 175 million and a median age of about 26 \cite{worldometers_bd_population2025}, there is enormous potential here. The education system serves about 40 million students across 130,000 schools \cite{moe2020}. Economic growth has been strong at six to seven percent annually, and the government is pushing digital initiatives \cite{a2i2021}. But the reality on the ground paints a different picture: dropout rates remain high, learning outcomes lag behind, and the distance between urban and rural schools is more visible than ever. According to recent national data, the average teacher–student ratio is approximately 1:34 \cite{banbeis_education_stats_2023}, electricity comes and goes, and internet connectivity is in best cases unreliable \cite{islam2019,alam2018}.

\begin{figure}[!t]
\centering
\includegraphics[width=0.9\columnwidth]{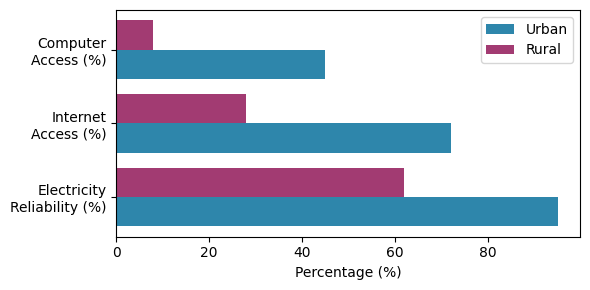}
\caption{Urban-rural access disparity in Bangladeshi schools showing significant gaps in electricity reliability, internet access, and student-computer ratios. Data highlights infrastructure challenges that existing educational technology frameworks fail to address.}
\label{fig:disparity}
\end{figure}

\begin{table}[!t]
\caption{Urban–Rural Disparities in School Infrastructure in Bangladesh}
\label{tab:education_snapshot}
\centering
\begin{tabular}{lccl}
\hline
\textbf{Metric} & \textbf{Urban} & \textbf{Rural} & \textbf{Source} \\
\hline
Student-teacher ratio & 35:1 & 52:1 & \cite{moe2020} \\
Computer access (\%) & 45 & 8 & \cite{alam2018} \\
Internet penetration (\%) & 72 & 28 & \cite{a2i2021} \\
Electricity reliability & 95\% & 62\% & \cite{islam2019} \\
(hours/day) & (22-24h) & (12-16h) & \\
\hline
\end{tabular}
\end{table}

When we visit many classrooms across Bangladesh, we often see students quietly listening while a teacher lectures from the front. There is little opportunity for hands-on learning, and this is a major challenge for subjects like mathematics and science that require active exploration. Our Table~\ref{tab:education_snapshot} outlines the sharp disparities between urban and rural schools, and Fig.~\ref{fig:disparity} visualizes how wide those gaps can be.

\subsection{Limitations of Existing Minecraft-in-Education Research}

Recent studies indicate that Minecraft can enhance engagement, spatial reasoning, and STEM learning; however, such outcomes often require stable electricity, reliable devices, and consistent teacher support \cite{slattery2023_minecraft_edu, baek2020_minecraft_implications}. In many low-resource settings such as rural Bangladesh, these assumptions frequently do not hold, undermining both access and learning potential. Moreover, most global studies do not fully account for socio-cultural and institutional constraints prevalent in developing countries—such as large class sizes, language and digital-literacy gaps, exam-oriented curricula, and minimal teacher training \cite{joshi2025_ict_barriers}. As a result, there is almost no rigorous work adapting Minecraft-based education to under-resourced, multilingual, exam-focused environments. Therefore, while Minecraft remains pedagogically promising, directly applying existing findings to contexts like Bangladesh without contextual adaptation is risky—highlighting the importance of context-aware, resource-sensitive frameworks such as ours.

\subsection{Scope and Positioning of This Paper}
This paper positions itself as a design-oriented, conceptual contribution rather than an empirical deployment study. Instead of reporting early pilot data which would be premature in a context with such complex infrastructural and pedagogical constraints, we provide an implementation-ready blueprint that integrates technical, curricular, interaction, and institutional dimensions.

\subsection{Socio-Cultural and Pedagogical Barriers in Bangladeshi Classrooms}

Deploying game-based learning in Bangladesh requires acknowledging several structural and cultural obstacles. Many rural and semi-urban schools lack reliable electricity, stable internet, or adequate devices, making game-based solutions difficult to implement in practice \cite{slattery2024_minecraft_eff, dileo2025_minecraft_review}. Teachers in under-resourced schools frequently have low ICT confidence and limited experience with interactive pedagogy, which complicates adoption without careful training and support \cite{hossain2023_tech_integration_bd}. In addition, an exam-oriented curriculum and strong pressure to cover syllabus content leave little room for exploratory or project-based learning, which makes game-based activities difficult to justify unless they are closely aligned with curriculum requirements. Socio-cultural factors such as language barriers, gender norms, and community skepticism toward games may further limit participation and sustained use.  

Our framework addresses these constraints through a combination of offline-ready deployment tiers, Bangla localization, low-cost hardware assumptions, and a structured teacher training program aiming for a realistic, context-aware path for implementation in Bangladesh’s diverse school settings.

\subsection{Contributions}

To the best of our knowledge, this is the first work to propose a 
systematic, context-aware framework for integrating Minecraft-based learning 
into a developing-country education system, explicitly designed for the 
constraints and socio-cultural realities of Bangladesh. Our main 
contributions are:

\begin{enumerate}
  \item A four-layer framework that jointly integrates infrastructure, 
  curriculum, interaction design, and institutional support—an approach not 
  explored in prior Minecraft-in-education research.
  
  \item A three-tier deployment model enabling meaningful participation 
  across heterogeneous school environments, including fully offline and 
  low-spec rural contexts typically overlooked in existing studies.
  
  \item A set of interaction and pedagogical design principles optimized for 
  first-time computer users, covering accessibility, linguistic support, and 
  culturally familiar metaphors.
  
  \item Eight curriculum-aligned and culturally grounded Minecraft worlds 
  localized in Bangla, demonstrating how national curriculum standards can be 
  operationalized in game-based environments.
  
  \item A practical evaluation framework with context-appropriate metrics, 
  including learning gains, usability, engagement, and sustainability 
  indicators suitable for low-resource settings.
\end{enumerate}

\section{Background and Related Work}

Game-based learning (GBL) has recently shown strong potential for improving engagement and higher-order thinking across subjects \cite{videnovik2023_gbl_cs}. Within this area, Minecraft Education Edition has emerged as a flexible environment for exploration, construction, and problem-solving. Empirical evidence continues to grow: a 2024 cluster-randomised trial reports measurable spatial-reasoning gains among primary students \cite{slattery2024_minecraft_eff}, while a 2025 structured review highlights consistent benefits in spatial cognition, conceptual understanding, motivation, and engagement, alongside challenges such as hardware demands and teacher preparation requirements \cite{dileo2025_minecraft_review}. Additional studies also demonstrate Minecraft’s value in STEM learning and spatial skills development \cite{carbonell2021_minecraft_spatial}.

However, most existing Minecraft-in-education research is situated in well-resourced school systems that assume stable electricity, modern devices, internet access, and digitally confident teachers. In contrast, evidence from South Asia and Bangladesh shows persistent barriers including irregular electricity, limited device availability, low ICT readiness, and exam-centric pedagogy \cite{munira2024_e_reading, ijcrt2023_elearning_bd}. These structural and socio-cultural realities are rarely considered when designing GBL interventions, which makes it difficult for existing models to work effectively in under-resourced settings.

As a result, no peer-reviewed work has adapted Minecraft-based learning for low-resource, multilingual, exam-oriented environments such as those in Bangladesh. Our work addresses this gap by proposing a context-aware, infrastructure-tiered, and Bangla-localized framework designed specifically for the pedagogical and infrastructural conditions of Bangladeshi schools.

\begin{table}[!t]
  \caption{Multi-layered framework architecture: four interconnected layers with responsibilities}
  \label{tab:framework_layers}
  \centering
  \begin{tabular}{p{0.32\linewidth} p{0.62\linewidth}}
    \hline
    \textbf{Layer} & \textbf{Responsibility / Function} \\
    \hline
    Institutional \& Training Layer & Policy alignment, teacher professional development, partnerships, open licensing. \\[3pt]
    Interaction \& Pedagogy Layer & Inclusive UI/UX design, adaptive tutorials, teacher dashboards, accessibility tools (keyboard-only controls, scalable text, etc.). \\[3pt]
    Content \& Curriculum Layer & Culturally grounded worlds, curriculum mapping, bilingual (Bangla–English) resources, context-relevant learning environments. \\[3pt]
    Infrastructure Layer & Tiered deployment (urban / semi-urban / rural), hardware provisioning, reliable power supply (solar / backup), and connectivity or offline-ready distribution. \\ 
    \hline
  \end{tabular}
\end{table}

\section{Proposed Framework Design}

We designed a framework that addresses infrastructure, pedagogy, content, and institutional support together. Each piece works independently but they are stronger when combined.

\subsection{System Architecture}

The system architecture consists of four interdependent layers, as shown in Table~\ref{tab:framework_layers}. Schools can implement what works for their situation without needing everything at once.

\subsubsection{Infrastructure Layer}

The infrastructure layer adopts a three-tier deployment model aligned with actual variability across Bangladeshi schools. \textbf{Tier~1} supports the 15\% of urban schools with stable electricity and broadband, enabling cloud-hosted multiplayer on standard desktops and laptops. \textbf{Tier~2} targets semi-urban schools (\(\approx 30\%\)) using local WiFi multiplayer, intermittent cloud syncing, and optional solar backup. \textbf{Tier~3} serves the remaining rural schools (\(\approx 55\%\)) through fully offline operation on low-spec or refurbished devices, turn-based shared gameplay, and USB/SD-based content distribution. This tiered design ensures functional access regardless of connectivity or hardware constraints.

\subsubsection{Content and Curriculum Layer}

\begin{figure}[!t]
\centering
\includegraphics[width=\columnwidth]{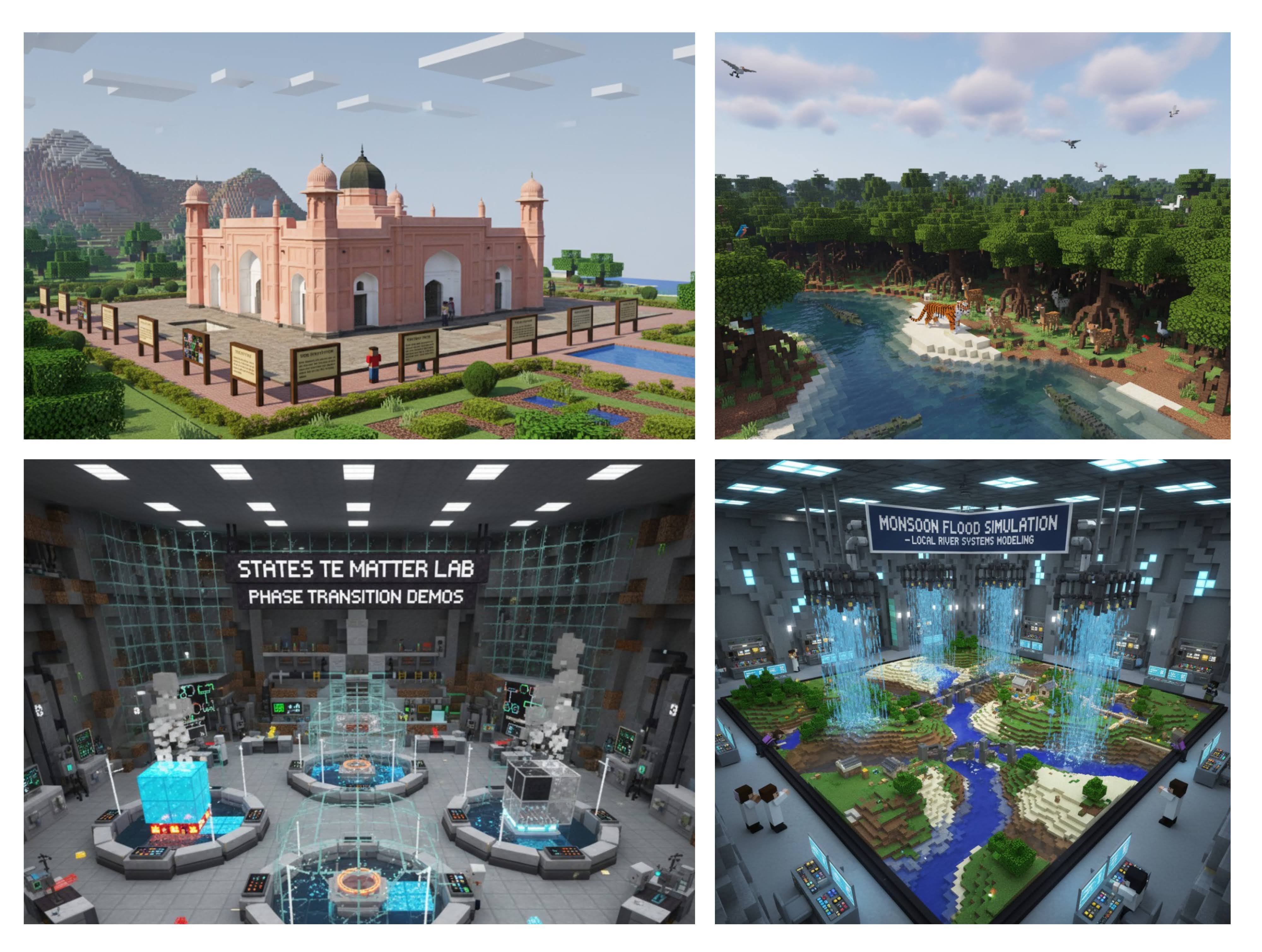}
\caption{Example curriculum-aligned Minecraft worlds used in the proposed framework: 
(a) Lalbagh Fort reconstruction with bilingual signboards supporting historical inquiry, 
(b) Sundarbans ecosystem featuring mangroves and wildlife for environmental studies, 
(c) States of Matter Lab demonstrating interactive phase transitions for science learning, 
(d) Monsoon Flood Simulation modeling local river systems and community-based mitigation strategies.}

\label{fig:minecraft_worlds}
\end{figure}

The content layer aligns activities with the national curriculum to ensure relevance to students’ formal learning objectives. We mapped Minecraft activities to National Curriculum and Textbook Board standards across science, math, social studies, and languages for grades three through twelve. Fig.~\ref{fig:minecraft_worlds} shows some of the pre-built worlds we are proposing.

\begin{table}[!t]
\caption{Detailed Curriculum-Aligned Minecraft Worlds}
\label{tab:curriculum_worlds}
\centering
\scriptsize
\begin{tabular}{p{1.3cm}p{0.8cm}p{2.6cm}p{2.4cm}}
\hline
\textbf{Subject} & \textbf{Grade} & \textbf{World Details} & \textbf{Learning Goals} \\
\hline
History & 9 & Lalbagh Fort with sandstone courtyard, bilingual lecterns & Mughal architecture, research skills \\
\hline
Environment & 10 & Sundarbans with mangrove trees, wildlife entities & Ecosystem dynamics, conservation \\
\hline
Civics & 11 & Shaheed Minar Liberation War monument & National identity, historical events \\
\hline
Science & 4 & Water Cycle with steam particles, redstone rain & Phase transitions, hydrological cycle \\
\hline
Science & 5 & States of Matter Lab with melting ice, cauldron boiler & Molecular behavior, temperature \\
\hline
Environment & 12 & Monsoon Flood with embankments, piston-gate release & Climate impacts, mitigation \\
\hline
Physics & 8 & Basic Circuits with redstone, daylight sensor & Electricity, renewable energy \\
\hline
Geography & 7 & Soil Erosion with hillside farm, terracing & Agriculture, land conservation \\
\hline
\end{tabular}
\end{table}

Table~\ref{tab:curriculum_worlds} summarizes the eight curriculum-aligned worlds. The history-focused environments (e.g., Lalbagh Fort, Shaheed Minar) provide bilingual signboards and authentic spatial layouts to support inquiry into Mughal architecture, the Liberation War, and national identity. Environmental and science worlds—such as the Sundarbans ecosystem, Water Cycle model, and States of Matter Lab—use Minecraft mechanics (e.g., steam effects, melting blocks, redstone triggers) to visualise key processes and enable guided experimentation. Applied STEM worlds, including the Monsoon Flood Simulation, Basic Circuits, and Soil Erosion scenarios, allow students to test mitigation strategies, construct simple electrical systems, and explore land-use concepts. All environments include full Bangla localization, voiced tutorials, and accessible in-world annotations to support diverse learners.

\subsubsection{Interaction and Pedagogy Layer}

The interface starts simple for young students and gradually unlocks features as they get comfortable. Early tutorials lock the camera angles and limit what is in the inventory so kids do not get overwhelmed. As students show they can handle it, more options open up. All tutorials have visual demonstrations with audio narration in both Bangla and English. We included colorblind-friendly color schemes, keyboard-only controls for schools without mice, captions for all audio, text that can scale up to 200 percent, and high-contrast modes.

\begin{figure}[!t]
\centering
\includegraphics[width=\columnwidth]{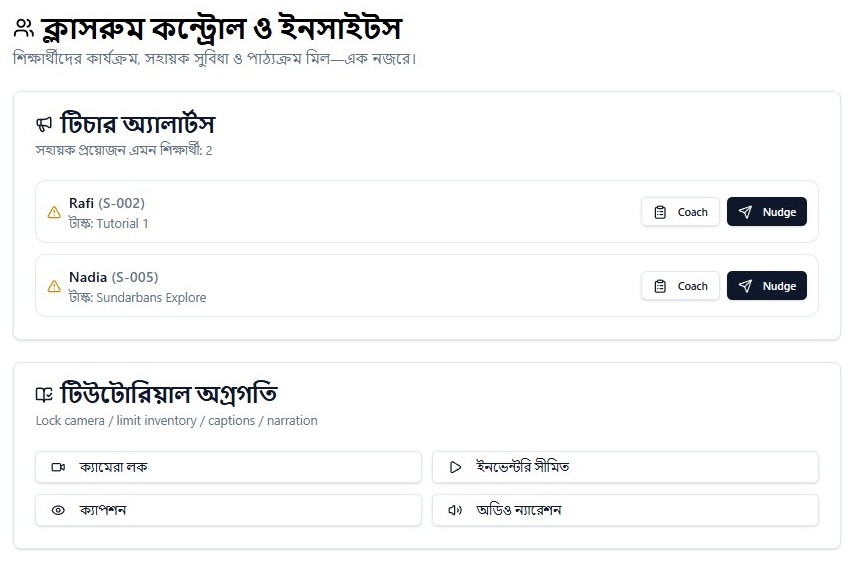}
\caption{Web-based teacher dashboard showing class monitoring tools, student activity status, and controls for camera lock, inventory limits, captions, and narration.
}
\label{fig:web}
\end{figure}

The teacher dashboard shown in Fig.~\ref{fig:web} provides a clear overview of class activity, displaying each student’s current task and status. Teachers can quickly check who is active, offer support through simple coaching prompts, and nudge students who may need guidance. The interface also includes classroom management controls such as locking the camera, limiting inventory, enabling captions, and toggling narration, allowing teachers to maintain a focused and accessible learning environment.

\subsubsection{Institutional and Training Layer}

Teachers need real training, not just a quick demo. We designed four modules: Module 1 covers basic digital literacy over 12 hours, Module 2 focuses on how to integrate this into teaching over 20 hours, Module 3 goes into advanced features over 16 hours, and Module 4 is an optional 24-hour deep dive into creating your own content. Beyond formal training, we are setting up video tutorial libraries in Bangla, peer mentoring networks so experienced teachers can help newcomers, online forums for troubleshooting, and regular webinars. On the policy side, we are emphasizing partnerships with education authorities, pilot programs that can inform larger rollouts, open licensing so content can be freely shared and adapted, and alignment with existing government digital initiatives.

\subsection{Feasibility of the Proposed Framework}

The framework is technically viable because it builds on deployment models already used in low-resource schools. Tier~3 uses an offline-first design with all worlds and assessments preloaded on local storage, eliminating dependence on continuous internet—an approach proven effective in prior ICT4D initiatives across South Asia and Africa \cite{burns2021_offline_tablets}. Reduced visual settings and turn-based interactions allow Minecraft Education Edition to run on low-spec or refurbished devices commonly available in Bangladeshi NGO programs \cite{minecraft_req_2025}.

Content distribution through USB/SD cards, lightweight LAN setups, and optional solar charging aligns with existing government digital-learning strategies. Progressive training and Bangla-localized interfaces further reduce adoption barriers for novice teachers and students. Together, these choices make the framework practical, implementable, and ready for pilot deployment. As Bangladesh advances its digital transformation, such scalable, resource-aware models can support more equitable STEM education nationwide.

\begin{table}[!t]
\caption{Layer 4: Teacher Training Curriculum}
\centering
\renewcommand{\arraystretch}{1.15}
\begin{tabular}{p{0.17\linewidth} p{0.08\linewidth} p{0.42\linewidth} p{0.17\linewidth}}
\hline
\textbf{Module} & \textbf{Hours} & \textbf{Core Competencies} & \textbf{Assessment} \\
\hline
\textbf{M1: Digital Basics} & 12 &
Device setup, file/USB workflows, offline/LAN modes, basic classroom safety & Skills check + micro-task \\
\textbf{M2: Pedagogy \& Alignment} & 20 &
Lesson planning with worlds, curriculum mapping (NCTB), activity scaffolds, bilingual supports & Lesson plan + demo \\
\textbf{M3: Orchestration \& A11y} & 16 &
Dashboard use, alerts/interventions, formative assessment, captions/HC/text-scale, keyboard-only flows & Live facilitation rubric \\
\textbf{M4: Authoring (Optional)} & 24 &
World editing, redstone scripting patterns, localization packs, template packaging \& open licensing & Publish a reusable world \\
\hline
\end{tabular}
\label{tab:layer4_training}
\end{table}

\section{HCI Design and Illustrative Scenario}

\subsection{Context-Specific HCI Principles}

To ensure usability in resource-constrained classrooms, we distilled
eight principles tailored to Bangladeshi schools
(Table~\ref{tab:hci_principles}).  These principles address both
first-time users (students and teachers) and pedagogical
effectiveness.

\begin{table}[!t]
\caption{Context-Specific HCI Principles for Minecraft-Based Learning}
\centering
\renewcommand{\arraystretch}{1.15}
\begin{tabular}{p{0.25\linewidth} p{0.65\linewidth}}
\hline
\textbf{Principle} & \textbf{Design Implication} \\
\hline
Progressive Complexity & Start with restricted camera angles and limited inventory; unlock features as learners gain confidence.\\
Culturally-Familiar Metaphors & Use farming, monsoon, and local architecture examples in tutorials rather than Western mining/crafting tropes.\\
Minimalist Interface & Hide non-essential menus; prefer icons over long text for beginners.\\
Linguistic Accessibility & Provide full Bangla localization with audio voice-over to support students with limited reading skills.\\
Gender-Inclusive Design & Emphasize collaborative tasks, offer diverse avatars, and avoid competitive ranking.\\
Disability Accommodation & Supply color-blind–friendly palettes, keyboard-only controls for mouse-less schools, adjustable text up to 200\%.\\
Economic Accessibility & Ensure all features operate on Tier-3 offline devices; no paywalls or mandatory internet connectivity.\\
Intrinsic Motivation & Leverage Minecraft’s natural curiosity-driven play instead of external points or grades.\\
\hline
\end{tabular}
\label{tab:hci_principles}
\end{table}

These principles guide all interface and world-building choices,
so that students in rural, semi-urban, and urban schools experience
a consistent and culturally relevant learning environment.


\subsection{Scenario: STEM Learning in a Rural Primary School}

\begin{figure}[!t]
\centering
\includegraphics[width=0.6\linewidth]{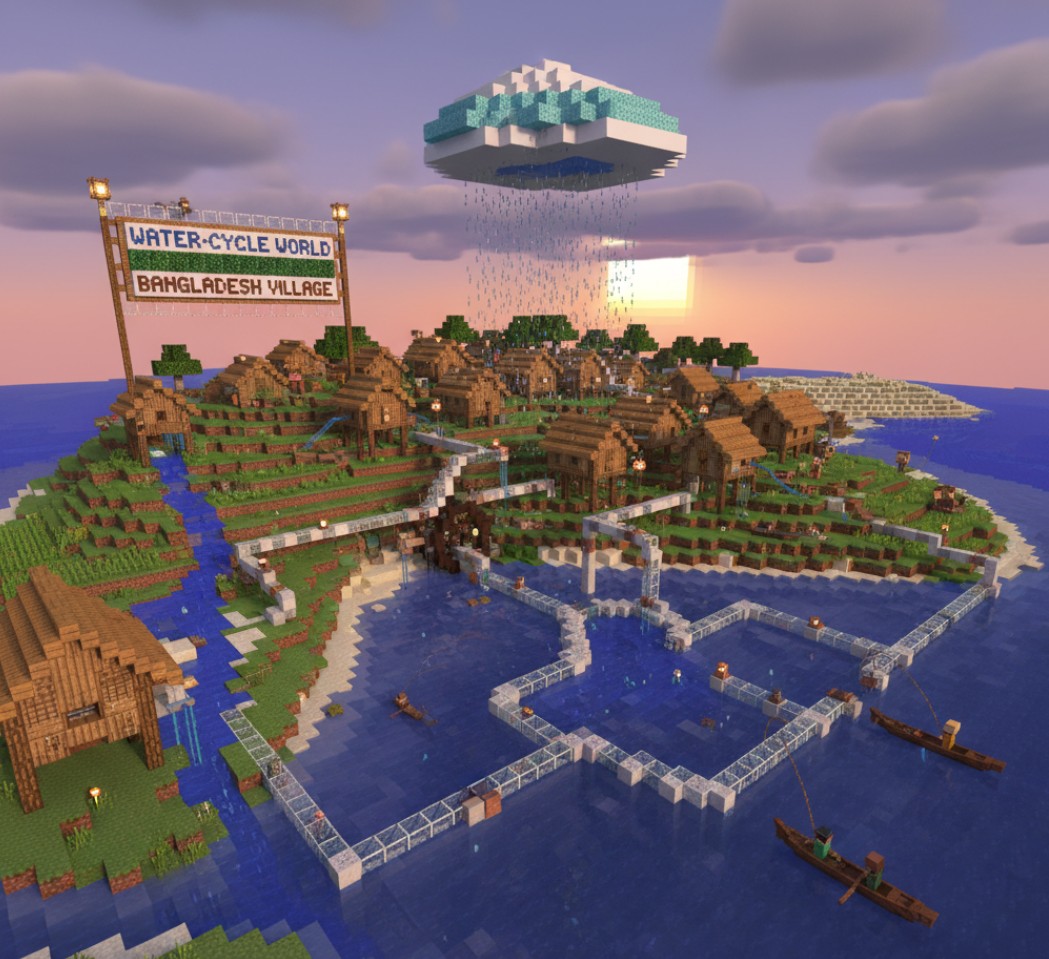}
\caption{Illustrative classroom scenario for the Water-Cycle lesson:
students explore phase changes and precipitation in a localized
Minecraft world while the teacher orchestrates discussion.}
\label{fig:scenario}
\end{figure}

Figure~\ref{fig:scenario} illustrates the lesson flow for a
Tier-3 deployment in a government primary school in
Mymensingh district.  The school has 180 pupils in Grades 3–5,
ten low-cost Android tablets, intermittent electricity,
and no internet connectivity.  A solar panel charges the tablets
and the teacher’s tablet acts as a local server for turn-based
multiplayer.

\begin{table}[htbp]
\caption{Lesson Flow in the Water-Cycle Scenario}
\centering
\renewcommand{\arraystretch}{1.1}
\begin{tabular}{p{0.23\linewidth} p{0.7\linewidth}}
\hline
\textbf{Stage} & \textbf{Activity and Learning Focus} \\
\hline
Tutorial (15 min) & Guided Bangla tutorial on block placement/removal inside the \emph{States of Matter Lab}. \\
Exploration (20 min) & Students try different heat sources, observe ice $\rightarrow$ water $\rightarrow$ steam with visual particle effects. \\
Guided Discovery (25 min) & Teacher prompts inquiry linking observations to seasonal monsoon patterns. \\
Application (30 min) & Groups manipulate the \emph{Water-Cycle World} to trigger evaporation, cloud formation, and rainfall over local terrain. \\
Group Presentation (10 min) & Teams explain their models and relate water flow to agricultural needs familiar to their families. \\
\hline
\end{tabular}
\label{tab:lesson_flow}
\end{table}

Evidence of understanding is captured in students’ saved Minecraft worlds and rubric-based evaluations: ability to model phase transitions, relate them to local contexts, and explain reasoning. Hands-on interaction makes abstract molecular concepts visible, eliminates the need for costly lab equipment,
and builds spatial reasoning and scientific inquiry skills.

\section{Evaluation Framework}

We need to know if this actually works, which means evaluating from multiple angles. 
\begin{figure}[htbp]
\centering
\includegraphics[width=0.9\linewidth]{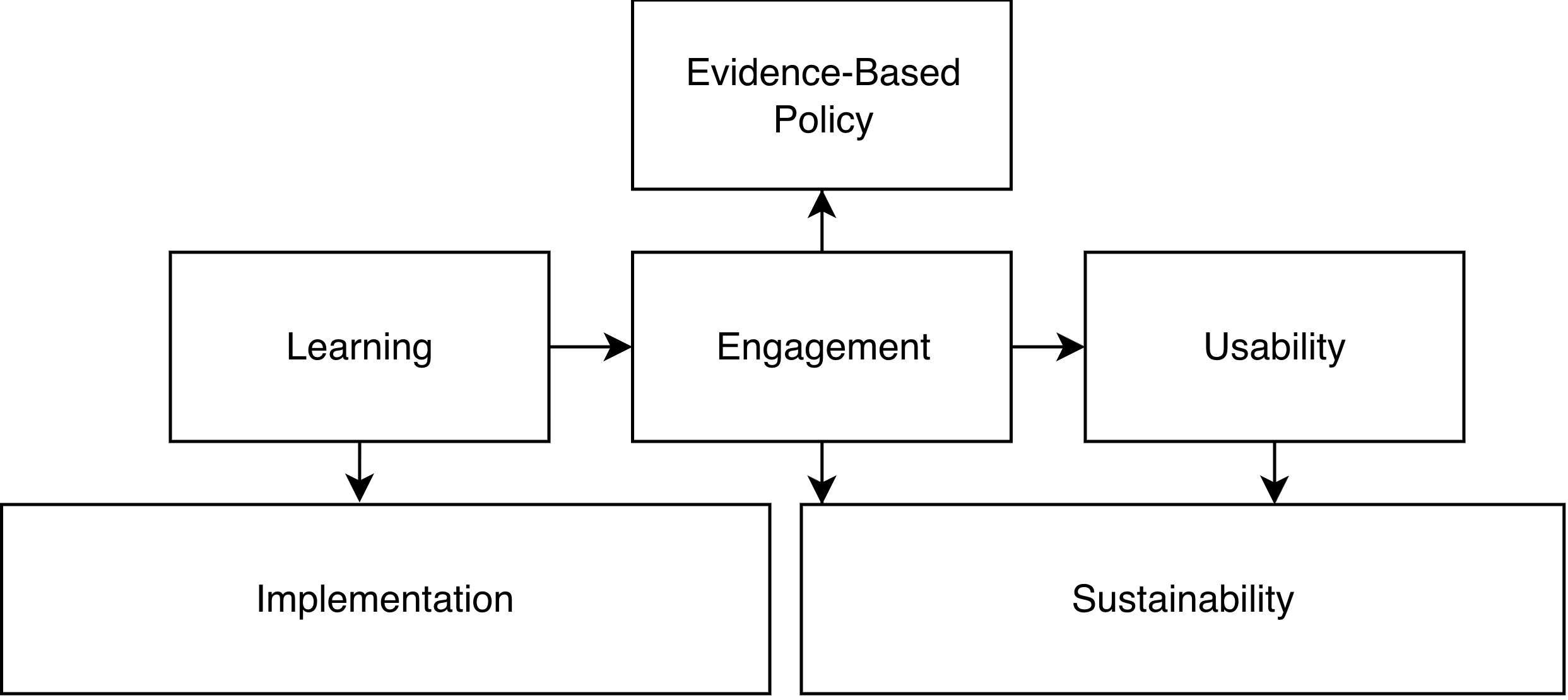}
\caption{Evaluation Pyramid illustrating how learning, engagement, and usability outcomes feed into implementation and sustainability, ultimately informing evidence-based policy.}

\label{fig:evaluation_pyramid}
\end{figure}

Fig.~\ref{fig:evaluation_pyramid} illustrates our multi-level evaluation,
linking learning outcomes to long-term sustainability and
evidence-based policy. We focus on three main domains:
\emph{learning}, assessed via pre-/post-tests, transfer tasks,
and sustained STEM interest; \emph{engagement and motivation},
tracked through validated Bangla-adapted surveys
(Intrinsic Motivation Inventory, Flow Short Scale) and in-game
analytics of building activity, session length, and collaboration;
and \emph{usability}, evaluated by think-aloud trials with
15–20 students per tier, success rates on common tasks,
and expert review of accessibility and cultural fit.

Implementation is monitored by interviewing teachers,
auditing infrastructure, estimating real costs, and tracking
whether teachers continue using the system and producing
their own worlds over time.  
Table~\ref{tab:evaluation_metrics} summarizes the target
benchmarks that define meaningful educational impact and
practical sustainability.

\begin{table}[t]
\caption{Evaluation Metrics at a Glance}
\label{tab:evaluation_metrics}
\centering
\small
\begin{tabular}{p{2cm}p{3cm}p{2.5cm}}
\hline
\textbf{Domain} & \textbf{Metric / Instrument} & \textbf{Target} \\
\hline
Learning & Pre–post test scores & +15\% gain \\
         & Transfer tasks      & 70\% mastery \\
         & STEM interest       & +20\% retention \\
\hline
Engagement & Intrinsic Motivation Inventory & $>$5.0 / 7.0 \\
           & Flow Short Scale               & $>$4.0 / 7.0 \\
           & Session duration               & $>$30 min/session \\
\hline
Usability  & System Usability Scale         & $>$70 \\
           & Task completion                & $>$85\% success \\
           & Think-aloud errors             & $<$3 per session \\
\hline
Sustainability & Continued usage            & $>$60\% at 1 yr \\
               & Teacher-authored content   & $>$3 worlds / term \\
               & Cost per student-hour      & $<$\$2 USD \\
\hline
\end{tabular}
\end{table}

\section{Discussion}

The proposed framework shows how Minecraft-based learning can extend beyond well-resourced classrooms by combining tiered infrastructure, localized content, and simplified interaction design, enabling hands-on, inquiry-driven learning even in schools with limited connectivity, low-spec hardware, or unreliable electricity. Such an approach aligns closely with national goals for digital literacy, STEM preparedness, and inclusive access under the Digital Bangladesh Vision. Nonetheless, challenges persist: infrastructure gaps, teacher ICT readiness, exam-focused pedagogy, gender norms, and community skepticism toward games continue to constrain adoption and require sustained training, localized content development, and institutional support. These realities highlight both the potential of the framework and the need for coordinated, long-term investment to ensure equitable and scalable implementation across diverse school contexts.

Policy-level commitment is therefore essential. Formal recognition of game-based learning, dedicated funding streams, integration into teacher education programs, and content-development support are critical for long-term sustainability. Equally important are peer-learning networks, district-level technical support, and mechanisms that allow teachers to adapt and share localized Minecraft worlds over time.

Finally, this work is inherently conceptual. No empirical classroom trials, cost analyses, or longitudinal data are yet available, and real-world conditions will undoubtedly introduce complexities not captured in the scenarios. Future research must therefore involve participatory design with teachers and students, pilot evaluations across different school tiers, and iterative refinement based on evidence.

\section{Conclusion}

This paper presents a design-oriented framework for integrating Minecraft into Bangladeshi classrooms, addressing infrastructural constraints, socio-cultural barriers, and pedagogical needs. While the framework remains conceptual and lacks empirical validation, it offers an implementation-ready blueprint grounded in offline-capable deployment tiers, localized content, and structured teacher support. Its primary limitation is the absence of field data, and real-world challenges—such as device availability, training demands, and community acceptance—remain to be tested. As next steps, we plan to collaborate with schools and education authorities to develop Bangla content, conduct pilot deployments, and refine the model based on empirical evidence. With continued research and partnership, this approach can contribute to future digital learning initiatives in Bangladesh and inform similar efforts in other resource-constrained contexts.

\bibliographystyle{IEEEtran}
\bibliography{references}

\end{document}